\input{aipcheck}

\documentclass[
    ,final            
  ]
  {aipproc}

\layoutstyle{6x9}
\usepackage{graphicx,natbib}

\begin{document}

\title{Short term aperiodic variability of X-ray binaries: its origin and implications}


\author{Mikhail Revnivtsev}{
  address={Max-Planck-Institute f\"ur Astrophysik, Garching bei M\"unchen,
              Germany \\ Space Research Institute, Russian Academy of Sciences,
               Moscow, Russia\\ Excellence Cluster Universe, Technische Universit\"at M\"unchen, Garching, Germany
}
}

\begin{abstract}
In this review I briefly describe the latest advances in 
studies of {\sl aperiodic} variability of accreting X-ray binaries and outline the
model which currently describe the majority of observational appearances 
of variability
of accreting sources in the best way. Then I concentrate on the case
of luminous accreting neutron star binaries (in the soft/high spectral state), 
where study of variability of X-ray emission of sources allowed us to 
resolve long standing problem of disentangling the contribution of
 accretion disk and boundary/spreading layer components to the time average
spectrum of sources. The obtained knowledge of the shape of the
spectrum of the boundary layer allowed us to make estimates of the
mass and radii of accreting neutron stars.
\end{abstract}

\maketitle


\section{Introduction}

Emission of X-ray binaries is not stable.
Already first obervations of X-ray binaries showed that their
emission variates on different time scales \cite[e.g.][]{oda71}. Some X-ray sources
appeared to be pulsars, which emission is modulated by rotation
of accreting neutron star \cite{giacconi71}, 
while variations of emission of some other 
sources were shown to be aperiodic \cite{oda71,rappaport71,terrell72}. 
Irrespective of the exact
origin of this variability it was immidiately realized that it brings us 
very important information about compact 
objects and environment around them \cite{oda71}, therefore a lot of 
efforts were devoted to study it.

In this review I will briefly describe the latest advances in our 
understanding
of aperiodic variability of accreting X-ray binaries and also will try to 
show how study of spectral variability properties of the accreting systems
helps us to disentangle different problems and to extract important 
information about compact objects.

\section{Shot noise and propagating flow models}

Very soon after discovery of aperiodic/flickering variability of X-ray binaries
(in particular, of Cyg X-1) it was proposed that such light curves might
be constructed out of a number of randomly occuring shots, which originate
in the region of main energy release in the accretion flow and have
some typical durations (i.e. ``shot noise model'', \cite{terrell72}). 
For quite a long
time this shot noise model was successfully applied to describe the
observed broad band variability of accreting X-ray sources 
\cite{nolan81,belloni90,lochner91}. However, 
collection of more and more observational facts about variability
of accreting sources (especially with the apperance of X-ray observatory
directly devoted to study the variations of X-ray emission of sources, Rossi 
X-ray Timing Explorer) led to the point where original shot noise model
should be modified in order to be more consistent with the data.

\begin{figure}
\includegraphics[width=.5\textwidth]{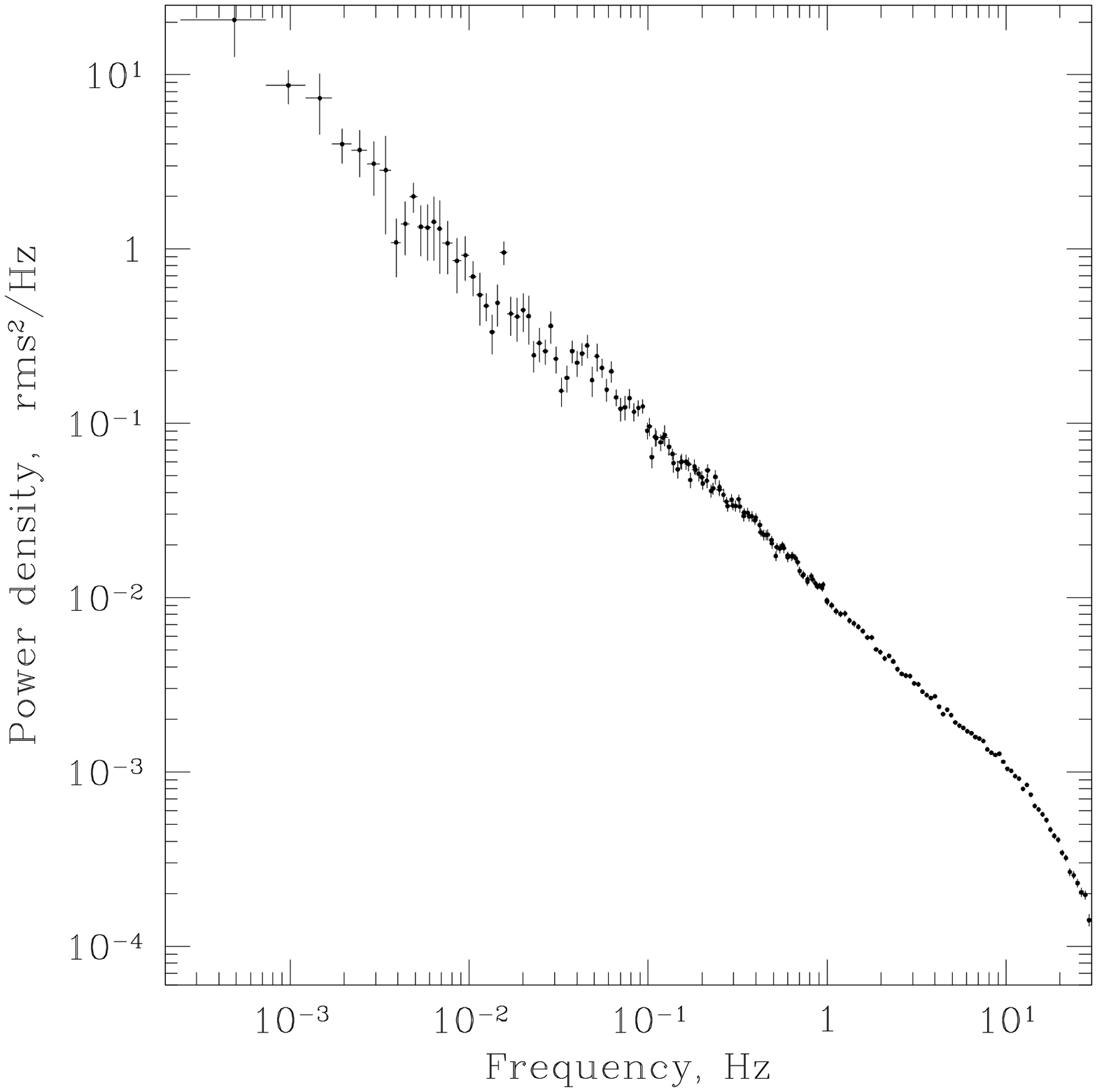}
\includegraphics[width=.5\textwidth]{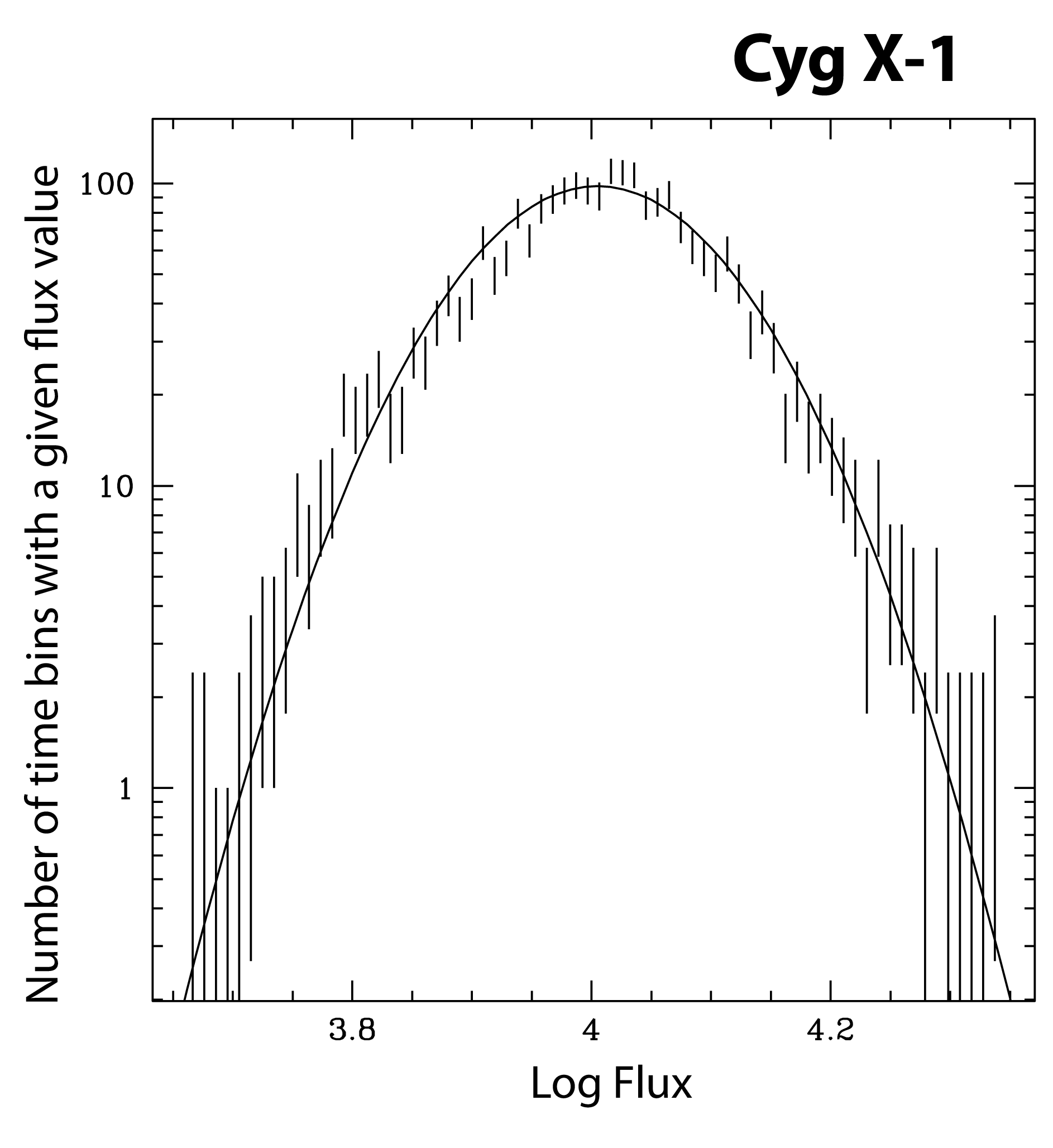}
  \caption{Problems of the original shot noise model. {\sl Left} -- power
spectrum of variability of the source (Cyg X-1 in the soft spectral state) 
on very wide range of time scales have a single slope.
 It is hard to imagine that all of
these times scales are produced within small region of main energy release
in the accretion flow. {\sl Right} -- distribution of values of
X-ray flux, constructed out of Cyg X-1 light curve, recorded with 1 sec time
bins. Clear lognormal distribution of the flux values (solid curve) 
can not be constructed
by additive summation of independent shots, implied by the original shot noise model }
\label{shotproblems}
\end{figure}

Among main problems of the original shot noise model one can mention two
issues:
\begin{enumerate}
\item Power spectra of some accreting sources (in particular, Cyg X-1 and 
the soft spectral state) can have the same power law shape over enormous 
range of frequencies, from tens of Hz down to $10^{-5}$ Hz and lower. 
The same slope of the power spectrum over this wide range of frequencies
suggests that variability at all these freqiencies originates as a result
of operation of the same physical
mechanism, but at the same time it is hard to imagine that 
all this huge range of
time scales can originate in the innermost region of the accretion flow, 
where dynamical time scales typically 5-8 orders of magnitude
smaller than the longest timescale of the observed variability 
\cite{churazov01} (see Fig.\ref{shotproblems}). 
\item The amplitude of the variability of the flux of accreting sources
scales linearly with the value of their time average flux \cite{lyutyi87,uttley01}, 
while in original shot noise model, which construct the light curve
via summation of randomly occuring independent shots, the amplitude of 
variability can not be directly related with the time average flux of 
the source. The distribution of values of X-ray flux 
typically have log-normal distribution rather then normal 
(see Fig.~\ref{shotproblems}), 
which should be expected if the flux of the source is constructed via
simple summation of independent shots.
\end{enumerate}

The solution of these problems was found in so called propagation flow 
models. 
In particular \citet{lyubarkii97} (some ideas were presented by
\citet{miyamoto88})
 considered the fluctuations of mass
accretion rate in the innermost region of the accretion flow associated 
with the fluctuations of the viscosity parameter $\alpha$ of the flow
at much larger radii. If the amplitude of fluctuations of $\alpha$ does
not depend on radius at which it occurs, the mass accretion rate 
at the innermost radius of the flow will have an $f^{-1}$ power density 
spectrum. Thus in this model the 
mass accretion rate is modulated at different distances from accreting object, 
but the observed (modulated) X--ray flux is coming from the
innermost region. The broad dynamic range of the variability time
scales in the model of \cite{lyubarkii97} is naturally provided by the broad
range of radii of accretion disk at which the viscosity is fluctuating. 

\begin{figure}
\vbox{
\hbox{
\includegraphics[width=0.4\textwidth]{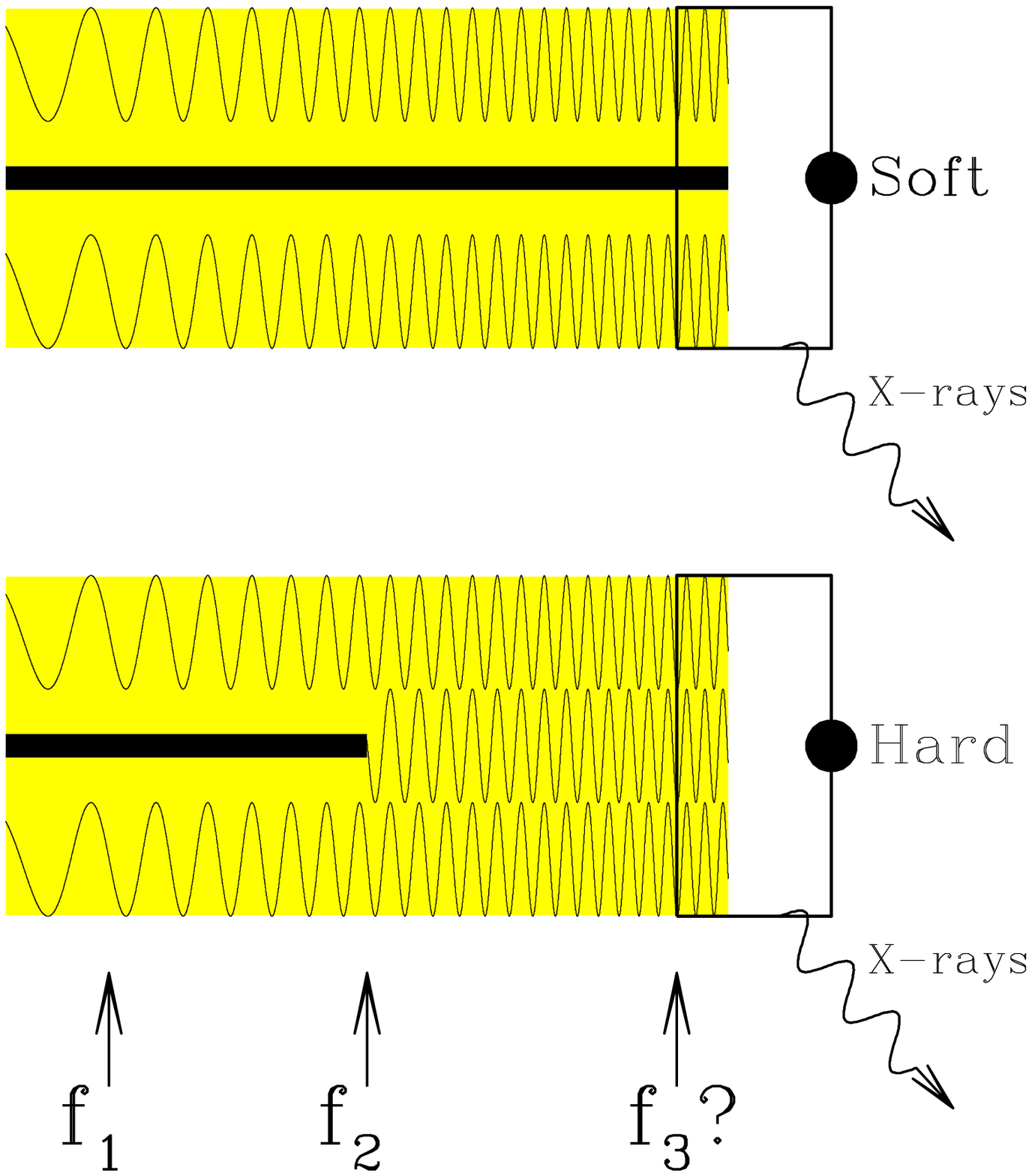}
\includegraphics[width=0.4\textwidth]{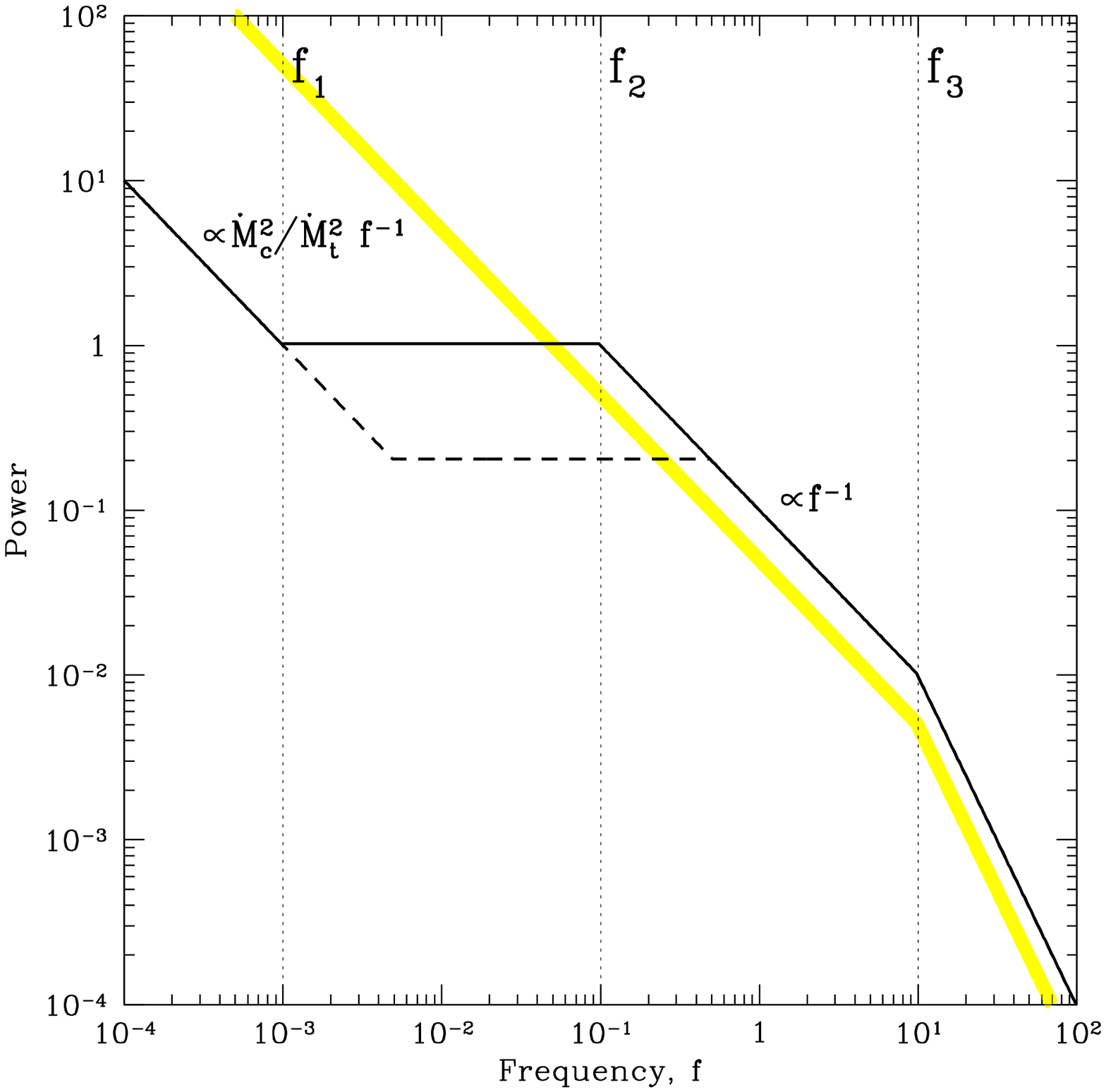}
}
\begin{center}
\includegraphics[width=0.4\textwidth]{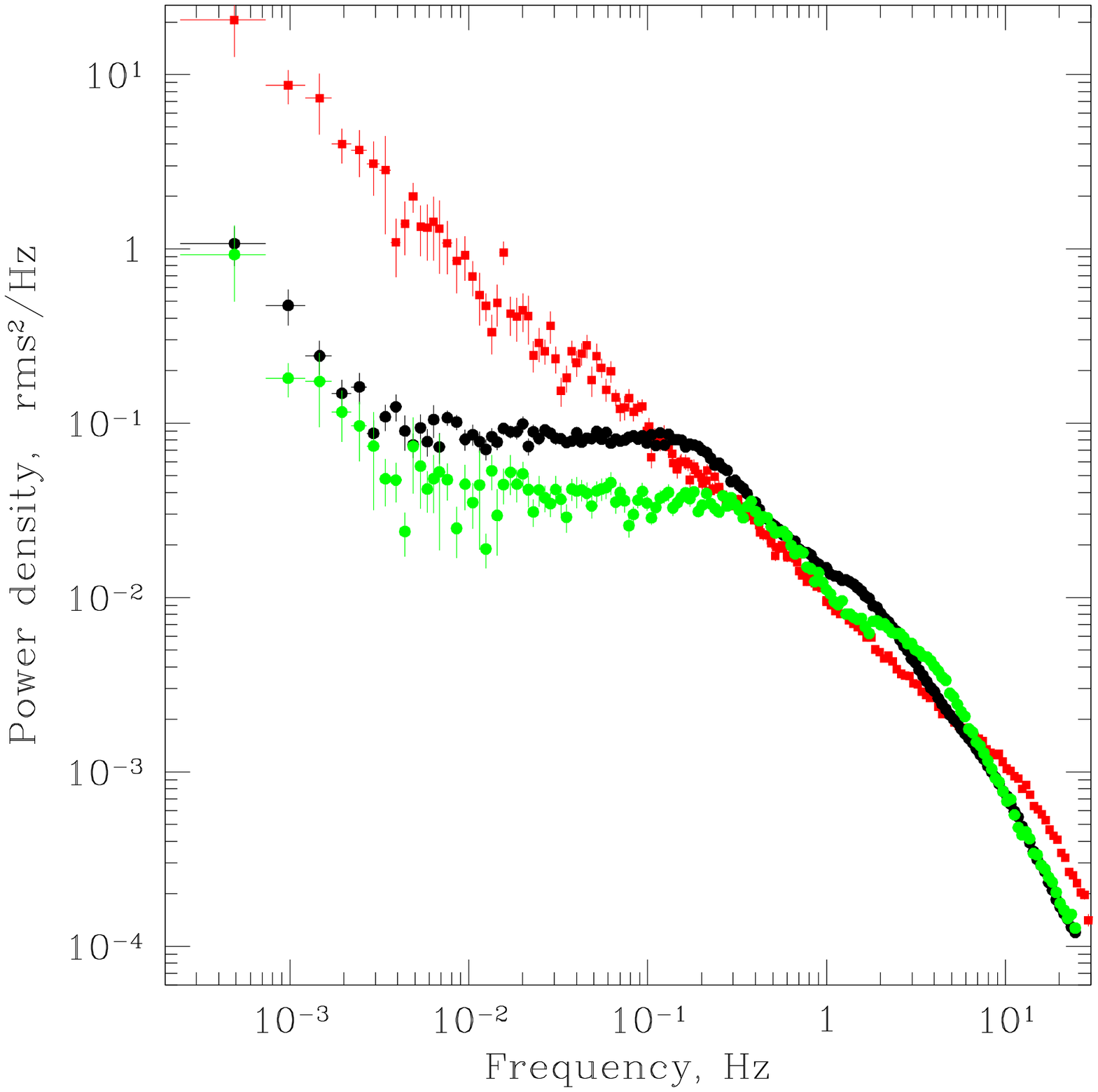}
\end{center}
}
\caption{{\sl Top--Left:} Sketch
of the assumed geometry for the soft and hard states 
of Cygnus X-1. The solid circle marks a position of a black hole. The
box shown by thin lines shows the area where most of the gravitational
energy is released and where most of the X--ray radiation is
emitted. The black ``slab'' shows the optically thick (geometrically
thin) accretion disk. In the soft state the inner edge of the disk is
close to the black hole, while in the hard state it is truncated far
from the energy release region. Sandwiching the disk
is an optically thin, geometrically thick corona (grey shaded regions),
extending in the radial direction up to a large distance from the
black hole. Oscillating curves show schematically that at different
radii the mass accretion rate in 
the corona is modulated on different time scales. This modulated
accretion flow reaches the innermost region and causes the
fluctuations of the observed X--ray flux over the broad range of the
time scales. 
{\sl Top--Right:} The overall shape of the power spectra expected in the simple
geometry adopted here. In the hard state (thick solid line) there are
three breaks ($f_1$,$f_2$,$f_3$ shown by thin vertical lines) in the
power spectrum. $f_2$ is the characteristic frequency in the optically
thin flow at the disk truncation radius. Anticipated changes in the
power density spectrum associated with the inward motion of the disk
truncation radius are shown by the dashed line. In the soft state the power
spectrum (thick grey line) is a power law up to $f_3$. 
{\sl  Bottom:} Typical power density spectra of Cygnus X-1 in
the hard (black and grey circles) and soft (squares) states. The
power spectra are constructed from the RXTE data in the 6--13 keV energy range.
From \cite{churazov01}.
\label{scheme}
}
\end{figure}

This model was elaborated and successfully applied to the data of Cyg X-1 in 
\cite{churazov01} (see Fig.\ref{scheme}). Subsequent application of this model to different 
objects showed that this model is able to reproduce the
shape of the power spectra of sources, dependence of their variability
amplitudes on the value of their flux, the distribution of
value of fluxes of sources, Fourier frequency dependent time and phase lags
and their dependence on energy \cite{churazov01,kotov01,uttley01,arevalo06}.

The main properties of variability of X-ray emission of
 accreting sources have following interpretation in the propagation flow model:

\begin{enumerate}
\item Emission of an accreting object emerges from  
the region of the main energy release in the accretion flow, but the
value of the mass accretion rate within this region at any given time 
is determined by its modulations inserted in the flow at broad range of 
distances from the compact 
object. If the structure of the accretion flow is the same at all distances, 
(as one might expect 
in the soft/high spectral state of source, see e.g. Fig.\ref{scheme}), one should expect to see
a power law shape of the power spectrum of the source variability from 
the shortest time scales, which can be created by instabilities in 
the innermost region of the accretion flow up to the longest time scales
for which the outermost regions of the accretion flow are responsible 
\cite{churazov01}
\item As the mass accretion rate in the innermost region of the accretion flow
is a result of {\sl multiplicative} summation of fluctuations, created
at different distances from the compact object, the distribution of the
source flux values naturally becomes lognormal and the amplitude of
short term variability is directly proportional to the current value of the
source flux \cite{uttley01,arevalo06}
\item At frequencies below which the outermost region of the accretion flow
can not produce variability the power spectrum of a source light 
curve becomes flat \cite{churazov01,gilfanov05}
\item Under assumption that the inner regions of the accretion flow emit 
slightly harder spectra of a power law shape, than the outer regions, the propagation flow model
naturally predicts the logarithmic dependence of the value of phase lag of light
curves in different energy channels \cite{kotov01} and the shape of the
phase lag dependence on Fourier frequency \cite{arevalo06}.  
\end{enumerate}

\section{Frequency resolved spectral analysis}

Time variability of X-ray flux of sources contains a lot of information
additional to commonly used time averaged characteristics.

 In particular, 
it may help to separate different components of the time averaged
energy spectrum in a model independent way. It is very easy to understand:
if, for example, the time averaged spectrum consists of two spectral 
components, one of which does not vary with time and another one varies
as a whole at some Fouier frequency, one can separate these two 
spectral components from each other without any a priori assumptions about 
their shapes and without any spectral modelling (see e.g early work on
this topic \cite{mitsuda84}).

The question about secure separation of spectral components is very important
in a number of cases. As an example, one can remind the modelling of the
fluorescent iron emission line in the spectra of accreting neutron stars and black 
holes (see e.g. \cite{reynolds03}). For such spectral decomposition the 
model independent approach is essential.

It is likely that the most advanced approach to study the combined 
spectral and timing 
information in the emission of sources is the Fourier frequency resolved 
spectral technique \cite{revnivtsev99}. In essence, it provides the
energy spectrum of a source variability in given Fourier frequency band 
in instrumental units, which can 
be unfolded using knowledge of the responce function of the instrument 
and standard packages like $XSPEC$ (see details in \cite{revnivtsev99,revnivtsev01}). 

\begin{figure}
\vbox{
\includegraphics[width=0.45\textwidth]{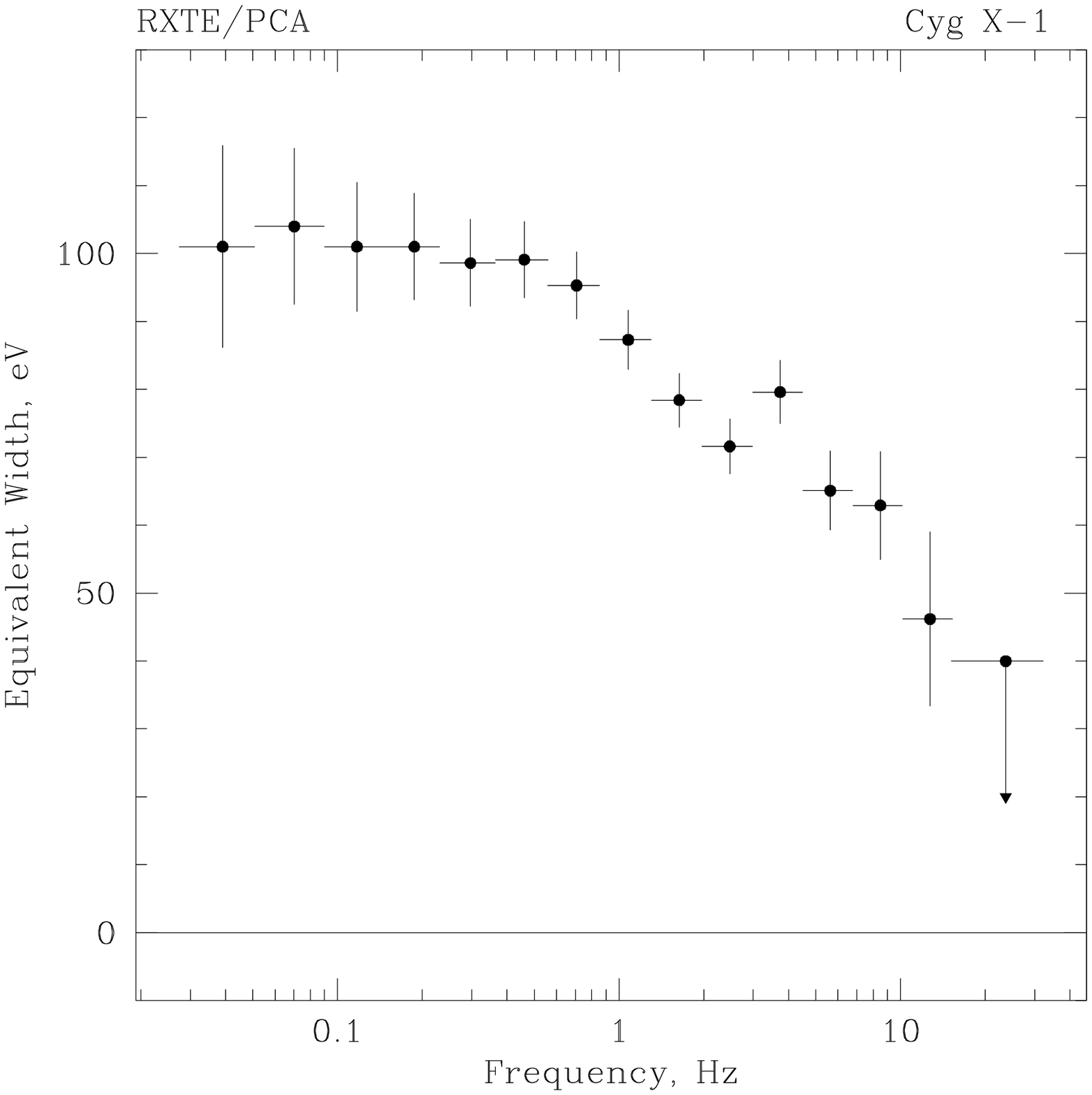}
\includegraphics[width=.45\textwidth]{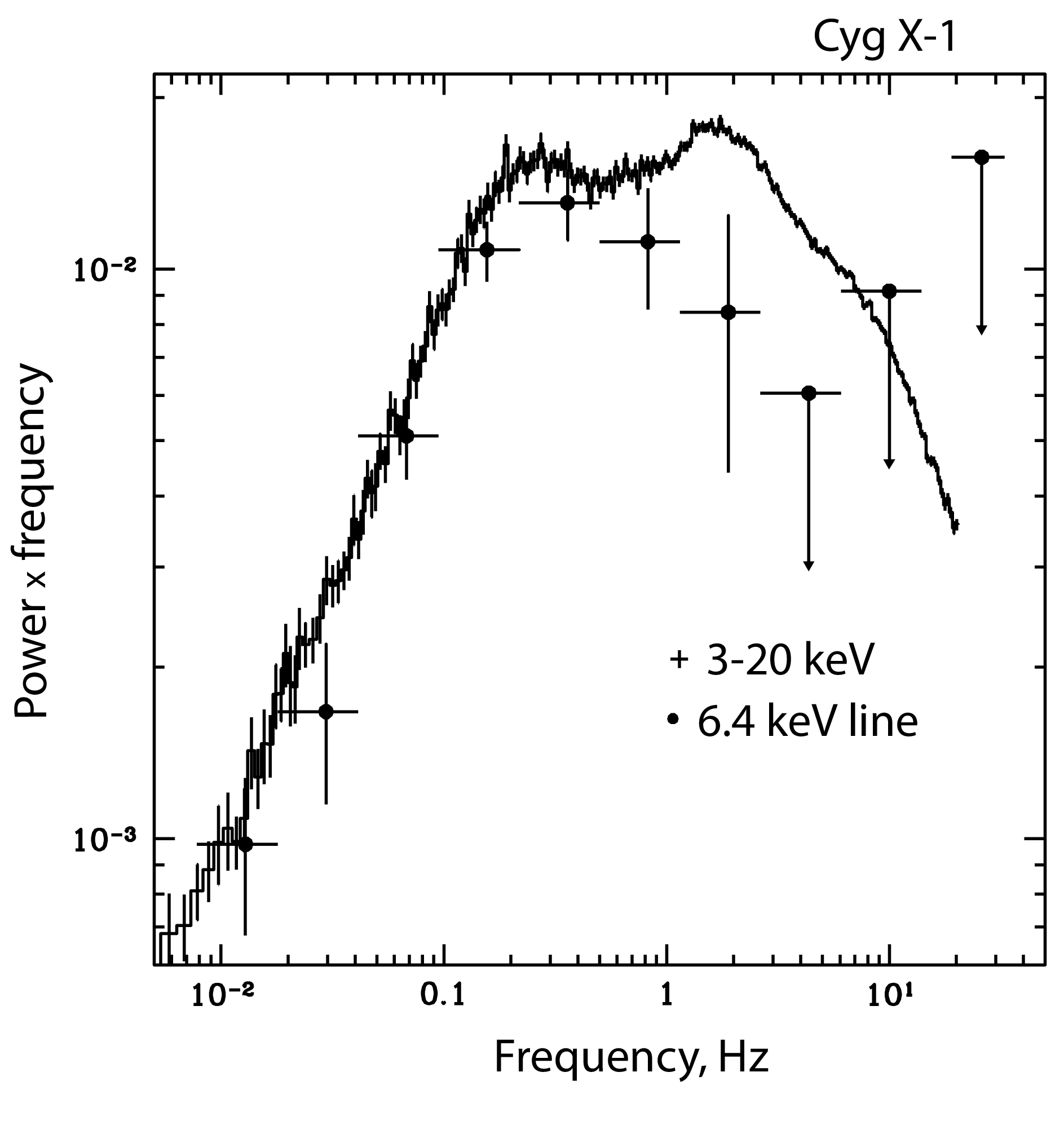}
}
  \caption{ {\sl Left} -- equivalent width of the fluorescent line
as a function of Fouier frequency at which the variabiluity spectrum is
constructed (see details in  \cite{revnivtsev99}). {\sl Right} -- the 
power spectra of the total flux of Cyg X-1 (3--20 keV) and the power
spectrum of the flux of the fluorescent line. The light curve of the
fluorescent line was constructed via direct decomposition of the energy 
spectrum of the source, accumulated in 1/64 sec time bins. }
\label{freq_res}
\end{figure}

For example, with the help of this method it was shown that the
fluorescent iron line ($\sim6.4$ keV) which is often visible in the spectrum
of accreting sources and originating, apparently, as result of
reprocessing of the hard X-ray emission of the central regions of the
accretion flow from relatively cold surrounding accretion disk, is much less variable
at high Fourier frequencies than the continuum emission \cite{revnivtsev99}.
Similar behavior of the variability of the iron line component in emission of
AGNs was revealed in \cite{reynolds00,papadakis05,papadakis07}.
This directly shows that these two spectral components (continuum and the
line) originate in geometrically distinct regions, and one of the
most probable explaination of such a behavior of the iron line is that
 the fastest innermost region do not produce the fluorescent line photons
due to further distance from the reflector/reprocessor
\cite{revnivtsev99}.

Fourier frequency resolved spectral technique of separation of
spectral components has significant advantages over other methods.
For example, in order to see the variability of some particular spectral 
component one can decompose the spectrum of the source, collected in
small time bins, construct the time history of the  flux of the distinct 
spectral component and then obtain the power spectrum of this light curve.
However, in this case (especially if the spectral model is not linear) 
the model fitting introduces significant additional noise in resulting
light curves and thus, the resulting power spectrum of the time history
of some component will be more noisy than if we would determine
the amplitudes of variability of the spectral component 
from Fourier frequency resolved energy spectra. 

On Fig. \ref{freq_res}
we present the dependence of the strength of the fluorescent iron 
line in the spectrum of Cyg X-1 (in hard spectral state) as a function of
Fourier frequency, obtained via Fourier frequency resolved spectral technique 
(left panel) and via construction of the power spectrum of variations
of the emission line component, which in tuen was obtained with the help of
fitting of the energy spectrum of the source
collected in 1/64 sec time bins. It is seen that results, obtained with 
these two methods are compatible, while the latter method gives much more noisy
results.

In the following part of the paper I will concentrate on one example where
application of the Fourier frequency resolved spectral technique 
allowed us to solve the long standing problem of separation of
spectral components in the soft/high state of neutron star binaries 
and allowed to make estimates of the mass and radii of accreting 
neutron stars.

\section{Emission of the neutron star boundary/spreading layer and masses
and radii of neutron stars}

Accreting neutron stars in low mass X-ray binaries (LMXB) are
among the most luminous compact X-ray sources in the Galaxy.
At least several of them have luminosities exceeding
$\sim {\rm few}\times 10^{38}$ erg/s, and they presumably
accrete matter at a level close to the critical Eddington accretion
rate. Early observations of these sources \citep[e.g.][]{toor70}
revealed rather soft  X-ray spectra, indicating that their X-ray
emission is predominantly formed in the optically thick media.
Similar to accreting black holes,  at lower X-ray luminosities (lower
mass accretion rates), $L_{\rm x}< 5\times 10^{36}$ erg/s,
neutron stars undergo a transition to the hard spectral state. 
The energy spectra in this state
point toward the low optical depth in the emission region.

In the soft spectral state, the commonly accepted picture of accretion
at values of the accretion rate that are not too extreme has
two main ingredients -- the  accretion disk (AD) and the boundary
layer (BL).  
While matter in the disk rotates with nearly Keplerian velocities, in
the boundary layer it decelerates down to the spin frequency of the
neutron star and settles onto its surface.
For the typical neutron star spin frequency ($<500-700$Hz), comparable
amounts of energy are released in  these two regions
\citep{sunyaev86,sibgatullin00}.  
This picture is based on rather obvious qualitative expectations, as
well as on more sophisticated theoretical considerations and numerical
modeling \citep{sunyaev86,inogamov99}.
This has been receiving, however, little direct observational
confirmation.  Due to the similarity of the spectra of the accretion disk
and boundary layer, the total spectrum has a smooth curved shape, which
is difficult to decompose into separate spectral components
\citep{mitsuda84,white88,disalvo02,done02}.   
This made it difficult to apply physically motivated spectral models to the
description of observed spectra of luminous neutron stars,
in spite of a very significant increase in the sensitivity of
X-ray instruments. A possible solution was suggested by early results
by \cite{mitsuda84}, who demonstrated the potential of using  
the combined spectral and variability information.

\begin{figure}
\includegraphics[width=.33\textwidth]{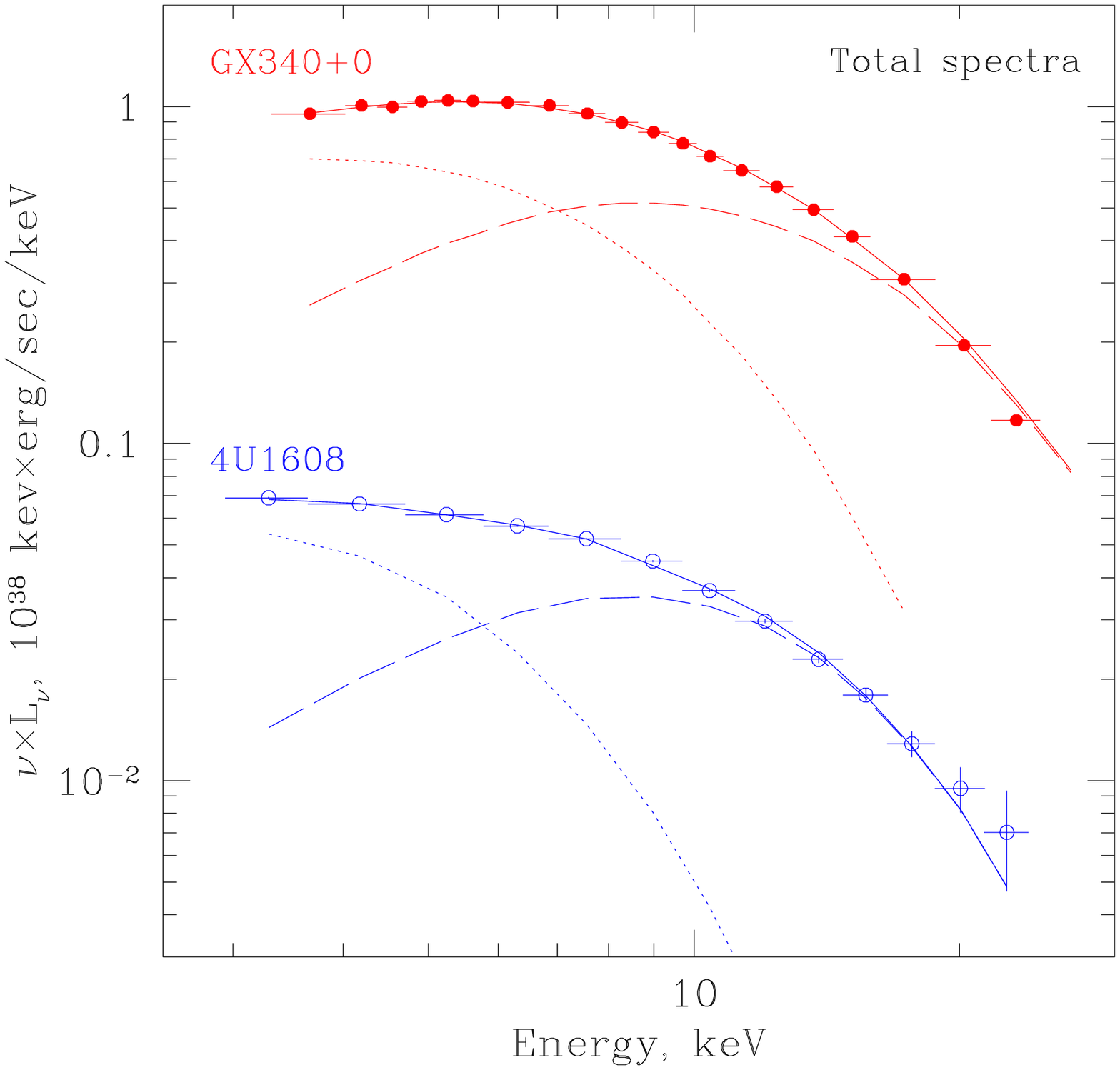}
\includegraphics[width=.33\textwidth]{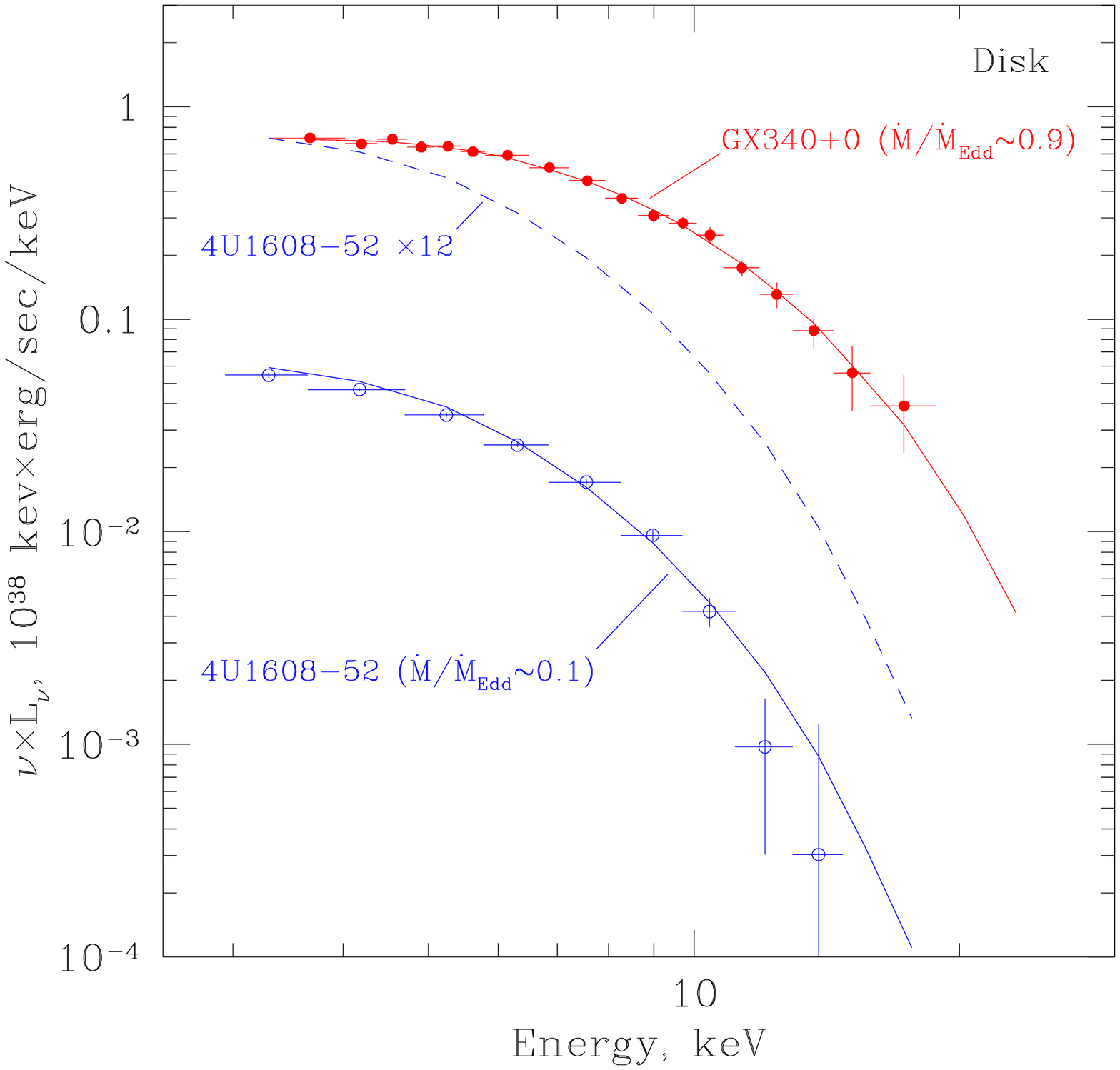}
\includegraphics[width=.33\textwidth]{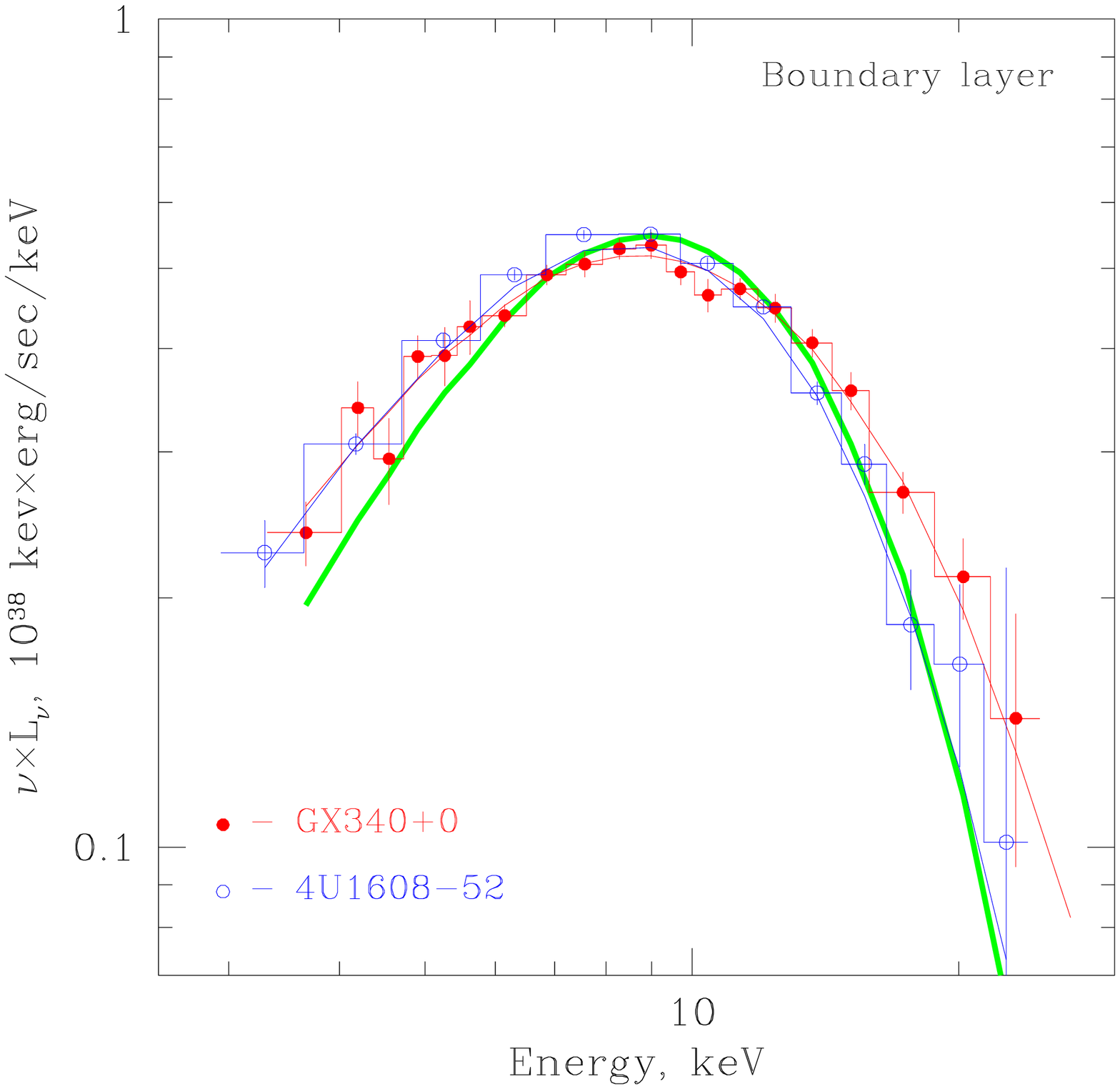}

\caption{{\sl Left} -- spectra of total emission of accreting neutron star binaries 
GX340+0 and 4U1608-52 decomposed into spectral components, originating
in accretion disk around the neutron star (dotted curve) and the 
boundary layer on its
surface (dashed curve). {\sl Center} -- spectra of the accretion disk in these 
two cases. Note the change of the temperature of the accretion disk. 
{\sl Right} -- spectra of the boundary layer in these two cases (the 
spectrum of the boundary layer of 4U1608-52 was scaled up to match those 
of GX340+0). It is clear 
that the shape of the BL spectrum virtually does not change in spite of
more than an order of magnitude difference in its luminosity.
 From \cite{gilfanov03}}
\label{nsbl}
\end{figure}

\begin{figure}
\includegraphics[width=.48\textwidth]{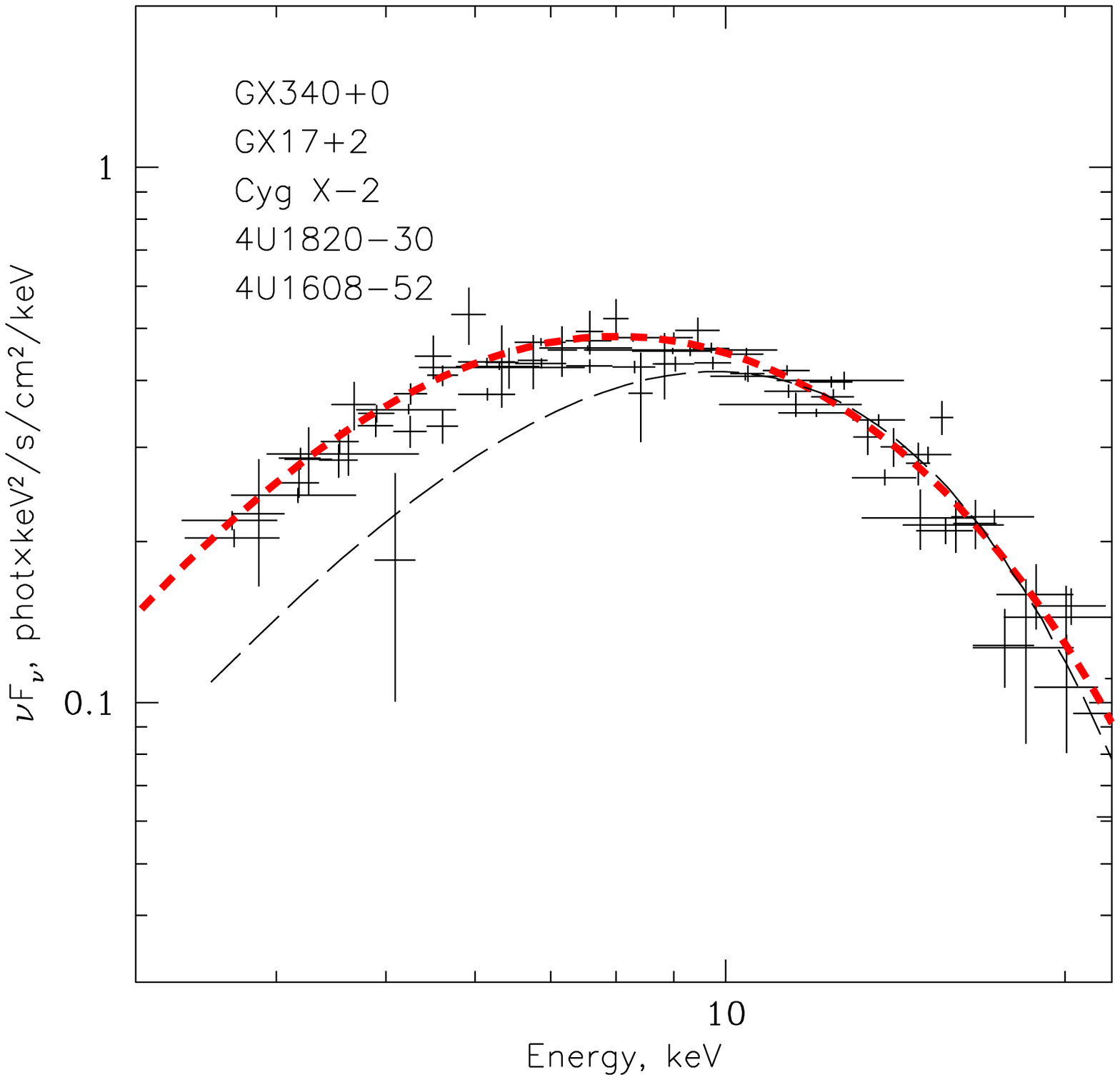}
\includegraphics[width=.5\textwidth]{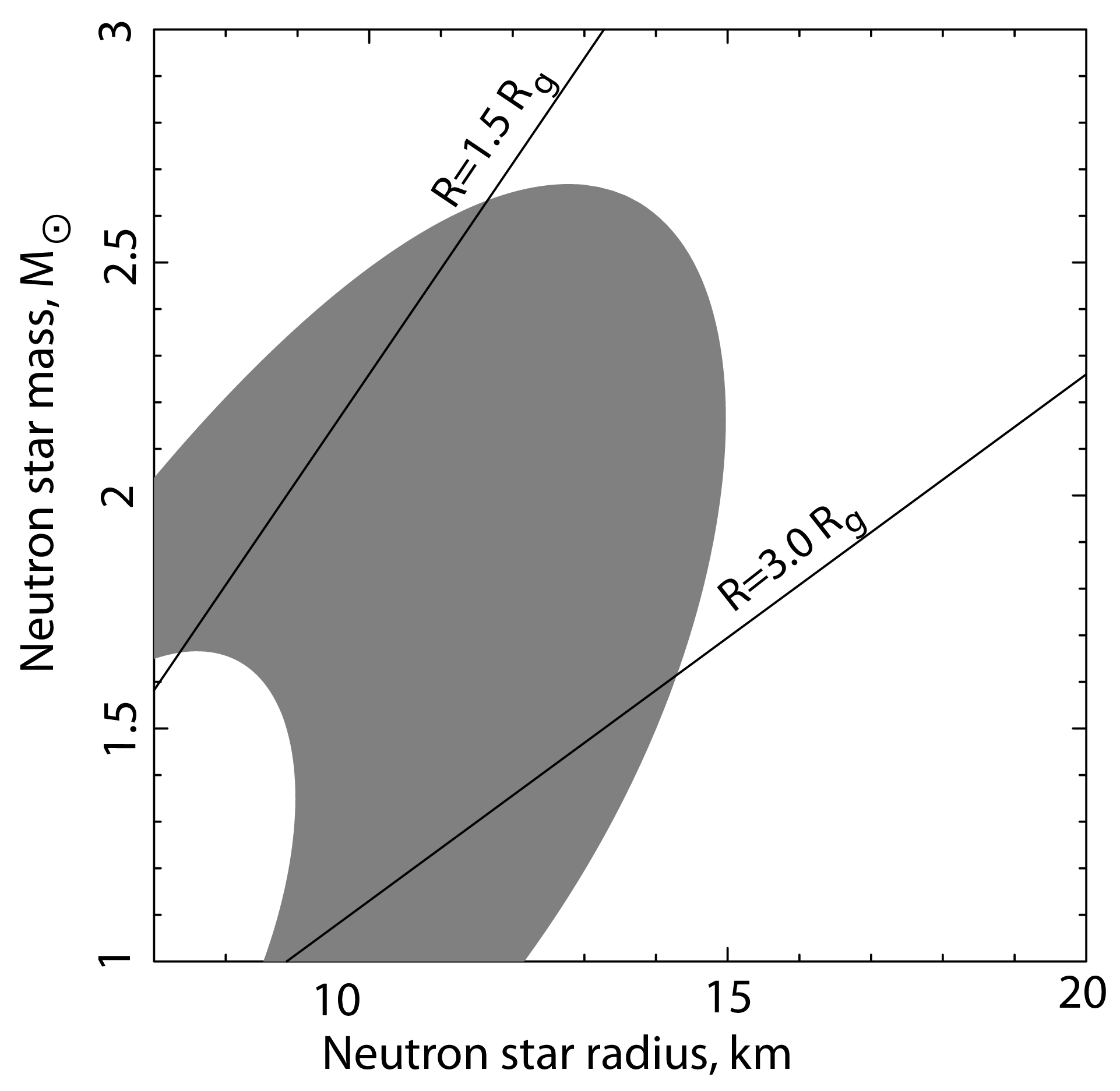}
  \caption{ {\sl Left} -- frequency resolved energy spectra (=spectra of the
boundary layer, see text) of a number of accreting neutron star binaries
in their soft spectral state. Thick dashed curve shows the best fit 
approximation of the spectra by a Comptonized emission model (see
\cite{revnivtsev06} for details). Thin dashed curve -- the black body
approximation of the spectra at energies $10-20$ keV. The color temperature
of the black body model is $kT\sim 2.4$ keV. {\sl Right} -- constrains on
masses and radii of accreting neutron stars from the measured value of the 
Eddington flux emerging from the photosphere of their boundary layers
(see text, details are presented in \cite{revnivtsev06})}
\label{mr_plot}
\end{figure}

Recently, \cite{gilfanov03} analyzed spectral variability in luminous
LMXBs and showed that in these sources aperiodic and quasi periodic
variability  on  $\sim$ sec -- msec time scales  is primarily caused
by variations in the luminosity of the boundary layer, while
the flux of the accretion disk remains almost perfectly stable at these 
frequencies (note that 
similar result about high stability of the X-ray flux of the accretion disk
in the case of accreting black hole Cyg X-1 in the soft spectral state 
was obtained previousely in \cite{churazov01}) .
Spectral shape of the boundary layer component remains nearly constant 
in the course of
the luminosity variations and is represented by the Fourier-frequency
resolved spectrum (see Fig.\ref{nsbl}). 
Moreover, in the considered  range $\dot{M}\sim
(0.1-1) \dot{M}_{\rm Edd}$ ($\dot{M}_{Edd}$ is the critical Eddington
mass accretion rate), it depends weakly on the global mass accretion
rate and in the limit $\dot{M}\sim \dot{M}_{\rm Edd}$ is close to the Wien
spectrum with $kT\sim 2.4$ keV.
In the work of \cite{revnivtsev06} it was shown that such a behavior 
is universal for all sources, for which the data allowed to construct 
frequency resolved energy spectra and the spectra of boundary layer
in all these cases are remarkably similar to each other in spite
of almost an order of magnitude difference in total luminosities of sources
(see Fig.\ref{mr_plot}, left panel).

Such behavior accords with the predictions of the model by
Inogamov \& Sunyaev (1999), namely, that at sufficiently high
accretion rates, $\dot{M}>0.1  \dot{M}_{Edd}$,  the boundary layer is
radiation-pressure dominated, and thus {\sl the local radiation flux 
is close to the critical Eddington value}. Increase of the mass accretion rate
leads to the increase of the emitting area of the BL, while
its vertical structure changes little \citep{inogamov99}.

It is important to notice that 
in this picture, the parameters of the BL emission can be used to
determine the value of the Eddington flux limit on the surface of the
neutron star. As the Eddington flux limit is uniquely determined by
the neutron star surface gravity and by the photospheric chemical
composition, the  neutron star mass and radius can be constrained.   

If the boundary layer emits true blackbody emission, the radiation
flux of the unit area was determined only by its temperature.
Therefore the observed shape of the BL spectrum, in particular the
bestfit blackbody temperature, could be used to determine the value
of the Eddington flux limit on the neutron star surface. This approach
has been utilized in the context of the Eddington limited
X-ray bursts \citep[e.g.][]{goldman79,marshall82,lewin93,
titarchuk02}. For fully ionized hydrogen atmosphere
$$
{\sigma T^4\over{c}} {\sigma_{\rm T} \over{ m_{\rm p}}} =
{G M (1-R_{\rm Sch}/R)^{3/2} \over{R^2}}
$$
where $\sigma$  is the Stefan-Boltzmann constant, $\sigma_{\rm T}$ 
is the
Thomson cross-section, $T$ is the blackbody temperature at infinity,
$m_{\rm p}$ is the proton mass, $M$ is the mass and $R$ is the radius of the
neutron star, $R_{\rm Sch}=2GM/c^2$ is the Schwarzschild radius of the
neutron star.
In addition, one would have to take into account that
the value of the Eddington flux is somewhat (by $\sim 10-20$\%)
reduced because of the action of the centrifugal force caused
by the rotation of the boundary layer
\citep{inogamov99}. Rotation of the neutron star at this point is 
not very important unless it is very high (rotational frequency 
$>$800-1000 Hz)

What is very important in our case -- are the scatterings in the atmosphere of the neutron
star,  therefore, the boundary layer spectrum will differ from the black
body (e.g. \cite{london86,lewin93}). The radiation transfer problem
in the atmosphere of the neutron star has been intensively
investigated, in particular in the context of X-ray bursts.
Numerical calculations  show that the effects
of scatterings can be approximately accounted for by introducing
the spectral hardening factor that relates the color and the
effective temperatures  of the emission
\citep{london86,titarchuk94,shimura_takahara95,ross96}.
Typical values of the hardening factor are about $\sim 1.7$.
We used this result in order to make simple estimates of the gravity
on the neutron star surface and to constrain its mass and radius.
In these calculations we assumed the color temperature of the boundary
layer emission $T=2.4 $ keV and considered the range of the
hardening factor values of 1.6-1.8.
We assumed that the centrifugal force reduces
the Eddington flux limit in comparison with the non-rotating boundary
layer by 20\%. We also took the finite height of the
boundary/spreading layer into account, which is about 1 km 
\citep{inogamov99}. The
calculations were performed for Schwarzschild geometry, hydrogen
atmosphere, and the result is shown in Fig.\ref{mr_plot}. The width of the
shaded region is defined by the assumed range of the values of the
hardening factor. Detailed modeling of the emergent spectrum of the
BL allows one to strongly diminish the size of uncertainty region
\cite{suleimanov06}.

\section{Conclusion}

Summarizing all of the above one can say that over last decade 
(with great help of high collecting area instruments aboard 
Rossi X-ray Timing Explorer) we have
achieved a big progress in understanding of aperiodic variability of 
accreting X-ray binaries. Existing models of propagating fluctuations 
allow us to 
explain the majority of observed properties of the aperiodic variability. 

Now with the help of Fouier frequency resolved spectral technique 
we are able to study combined spectral and timing variability of the
accreting binaries, which proved to be very efficient method of disentangling
the contribution of geometrically different regions to the total
time averaged spectrum of sources. Application of this method already 
allowed us to obtain separately the spectrum of the boundary/spreading
layer at the surface of accreting neutron stars, thus emphasizing the
importance of the surface effects in the case of accretion flow around
 neutron star binaries. Our findings support the theoretocal
predictions that the boundary/spreading layer at the neutron star 
surface at high mass accretim rates should be radiation pressure dominated. 
This in turn 
allowed us to obtain important constrains on masses and radii of neutron 
stars. We hope that with the appearance of new timing instruments on 
next generation of observatories (e.g. XEUS/HTRS, see \cite{barret06}), 
we might probe even stronger gravity in the innermost regions of the
accretion flow and variable phenomena on neutron stars surface under
extreme conditions of high matter density and radiation pressure dominated
environment.


\begin{theacknowledgments}
This work was supported by DFG-Schwerpunktprogramme
(SPP 1177), grants CH389/3-2, RFFI~07-02-01051,
07-02-00961, NSH-5579.2008.2 and by program of Presidium of 
Russian Academy of Sciences ``Formation and evolution of stars and galaxies''
\end{theacknowledgments}



\bibliographystyle{aipproc}   

\bibliography{sample}

\IfFileExists{\jobname.bbl}{}
 {\typeout{}
  \typeout{******************************************}
  \typeout{** Please run "bibtex \jobname" to optain}
  \typeout{** the bibliography and then re-run LaTeX}
  \typeout{** twice to fix the references!}
  \typeout{******************************************}
  \typeout{}
 }

\end{document}